\documentclass[conference]{IEEEtran}
\IEEEoverridecommandlockouts
\usepackage{cite}
\usepackage{amsmath,amssymb,amsfonts}
\usepackage{algorithmic}
\usepackage{graphicx}
\usepackage{textcomp}
\usepackage{amsmath}
\usepackage{bm}
\usepackage{algorithm}
\usepackage{color}
\usepackage{cite}
\usepackage{amssymb}
\usepackage{multirow}
\usepackage{epsfig}
\usepackage{framed}
\usepackage{stfloats}
\usepackage{lettrine}
\usepackage{mathtools}
\usepackage{array}
\usepackage{slashbox}
\usepackage{tabu}
\usepackage{caption}
\usepackage{subcaption}
\usepackage{nicefrac}
\usepackage{algorithm}
\usepackage{algorithmic}
\usepackage{makecell}
\usepackage{mathtools}

\def\BibTeX{{\rm B\kern-.05em{\sc i\kern-.025em b}\kern-.08em
    T\kern-.1667em\lower.7ex\hbox{E}\kern-.125emX}}
\begin{document}

\title{A Novel Self-Organizing PID Approach for Controlling Mobile Robot Locomotion
\thanks{The first three authors contribute equally. This work is funded by Innovate UK under the project ``Collaborative Technology Hardened for Underwater and Littoral Hazardous Environments” (Ref. 104061). * Corresponding Author: P. Angelov}
}

\author{\IEEEauthorblockN{Xiaowei Gu\textsuperscript{1,2},\textit{Member, IEEE}, Muhammad Aurangzeb Khan\textsuperscript{1,2}, Plamen Angelov\textsuperscript{1,2,*}, \textit{Fellow, IEEE}, \\  Bikash Tiwary\textsuperscript{1}, Elnaz Shafipour Yourdshah\textsuperscript{1,2} and Zhao-Xu Yang\textsuperscript{3}}
\IEEEauthorblockA{\textit{\textsuperscript{1}School of Computing and Communications,Lancaster University, UK} \\
\textit{\textsuperscript{2}Lancaster Intelligent, Robotic and Autonomous Systems Centre (LIRA), Lancaster University, UK} \\
\textit{\textsuperscript{3}State Key Laboratory for Strength and Vibration of Mechanical Structures,}\\
\textit{ Shaanxi Key Laboratory of Environment and Control for Flight Vehicle,}\\
\textit{School of Aerospace, Xi'an Jiaotong University, Xi'an, People's Republic of China}\\
* Email: p.angelov@lancaster.ac.uk}\\
}

\maketitle

\begin{abstract}
A novel self-organizing fuzzy proportional–integral–derivative (SOF-PID) control system is proposed in this paper. 
The proposed system consists of a pair of control and reference models, both of which are implemented by a first-order autonomous learning multiple model (ALMMo) neuro-fuzzy system. 
The SOF-PID controller self-organizes and self-updates the structures and meta-parameters of both the control and reference models during the control process ``on the fly”. 
This gives the SOF-PID control system the capability of quickly adapting to entirely new operating environments without a full re-training.
Moreover, the SOF-PID control system is free from user- and problem-specific parameters, and the uniform stability of the SOF-PID control system is theoretically guaranteed. 
Simulations and real-world experiments with mobile robots demonstrate the effectiveness and validity of the proposed SOF-PID control system.
\end{abstract}

\begin{IEEEkeywords}
 ALMMo neuro fuzzy system, PID controller, self-organizing system
\end{IEEEkeywords}

\section{Introduction}
Since being firstly introduced eighty years ago, proportional–integral–derivative (PID) controllers have been extensively used in industry and defense \cite{Singh2019,Marin2018,Gonzalez2018}
for automation and process control thanks to their merits of, low costs, inexpensive maintenance, simplicity in structure and capability for accurate control. 
More importantly, the control parameters of the PID controllers have clear physical meaning \cite{Bennett1993}, they are explainable and interpretable by human experts. 
The fundamentals of conventional PID controllers were summarized in \cite{Astrom1993}. \par

However, the well-known drawbacks of the PID controllers are: 1) there are some strong recommendations and algorithm regarding PID tuning such as linear stability techniques \cite{Ziegler1942,Silva2003}; 
2) the performance is fragile for nonlinear, complex and vague systems that have no precise mathematical models \cite{Tang2001}; 3) the performance is fragile for new data patterns. 
The empirical approaches for parameter-tuning usually can ensure a satisfactory performance of the PID controller, 
but it requires heavy involvements of human expertise and a good understanding of the problem \cite{Ziegler1942}. 
To address the second drawback, various types of modifications have been made,
 self-tuning adaptive PID controllers \cite{Astrom1992,Rosales2019} and fuzzy PID controllers \cite{Tang2001,Lapa2018,Dettori2018,Shao1988,Ferdaus2018,Angelov2005,Angelov2006} have been developed and applied widely. \par

The architecture of adaptive controllers are pre-fixed during the design stage, and the self-tuning schemes concerns 
adjusting control parameters only \cite{Gonzalez2018,Lapa2018,Dettori2018,Eltag2019,Huang2000}. Meanwhile, the vast majority of existing PID control models 
lack self-organizing structure that can be adaptive to entirely new environments. 
Therefore, in many real-world applications, e.g., autonomous driving and mobile robots, the performance of a well-tuned PID controller 
can significantly deteriorate when new situations appear. In other words, the adaptive (fuzzy) PID controllers are 
capable of dealing with the shifts in the data pattern, but are unable to handling the drifts \cite{Lughofer2011}. More recently, self-evolving
 fuzzy rule-based controller \cite{Angelov2004,Costa2004} was introduced, which is able to self-update the system architecture during the control process. 
In this paper, we propose to improve the evolving mechanisms of these self-evolving algorithms \cite{Angelov2018}. \par

In this paper, we propose a self-organizing fuzzy PID (SOF-PID) control system. The SOF-PID control system is composed of two parts:
 i) reference model and ii) control model. Both models are implemented as autonomous learning multi-model (ALMMo) neuro-fuzzy systems \cite{Angelov2018}.
 A control system consisting of two models was firstly introduced in \cite{Psaltis1988}, where two neural networks are involved serving as the respective feedforward and feedback controllers. 
This type of structure inherits the advantages of both, feedforward and feedback controllers, and, generally, results in a better tracking performance \cite{Fan2019}
 A number of control systems with similar architectures were introduced afterwards.
 \cite{Zhao1997} presents a generic algorithm-based feedforward-feedback multiple-input–multiple-output (MIMO) control system pressurized water reactor power plant.
 In \cite{Chiu2006}, a feedforward-feedback fuzzy adaptive controller consisting of a pair of a self-adjusting feedforward fuzzy model and a pre-fixed error-feedback 
model is designed for multiple-input–multiple-output uncertain nonlinear systems. A temperature controller for of functionally graded plates is proposed in \cite{GolbaharHaghigh2010}, 
which combines a feedforward inverse control model with a proportional–derivative (PD) controller for disturbance/noise attenuation. 
A robust evolving cloud-based controller is introduced in \cite{Andonovski2016}. This approach is composed of an AnYa type fuzzy rule-based system \cite{Angelov2012a,Angelov2012b}, 
which is used for heat-exchanger plant control, and a reference model, which is for producing the desired trajectory. 
Other successful implementation for real-world problems of feedforward-feedback control systems include, but not limited to sludge process pilot plant \cite{Vrecko2006}; 
atomic force microscopes \cite{Pao2007}; solid oxide fuel cell system \cite{Vrecko2018}; piezoelectric-driven mechanism \cite{Fan2019}, etc.
 Nonetheless, the system structures of most existing approaches are designed by human experts based on prior knowledge of the problems. 
The control systems can be partially adjusted only, while the rest parts have been pre-fixed. Although, a properly designed control system 
usually demonstrates excellent performance under the experimental scenarios, the system structure and parameters can be less meaningful, 
and the performance will deteriorate when the operating environment is changed significantly.\par

ALMMo neuro-fuzzy system \cite{Angelov2018} is a new type of first-order multi-model system based on AnYa type (neuro-)fuzzy systems \cite{Angelov2012a,Angelov2012b}.
Comparing with other more widely used types of (neuro-)fuzzy systems, e.g. Zadeh-Mamdani type \cite{Mamdani1975}, Takagi-Sugeno type \cite{Takagi1985}, 
AnYa type models simplify the design process by reducing it to the choice of prototypes only, 
which are the most representative data samples and can be identified in a fully data-driven, nonparametric manner.
 This simplification significantly reduces the involvement of human expertise and also lifts the requirement of prior knowledge of the problems for system identification. 
The system structure of the ALMMo system is built upon non-parametric data clouds identified from streaming data in an autonomous,
self-organizing and transparent manner without imposing any data generation models pre-defined in advance. 
The meta-parameters of the system are updated accordingly during the identification process without relying on any
 user- and problem-specific parameters. ALMMo system is a generic approach for nonlinear, nonstationary problem approximation,
 and it has demonstrated its strong performance on various benchmark problems \cite{Angelov2018} and successfully applied on real-world problems
 including stock market index prediction and foreign currency exchange rate prediction \cite{Gu2019,Yong2018}. \par

Employing the ALMMo neuro-fuzzy system as the learning engine, both, the control and reference models of the SOF-PID control system
 are able to learn from data streams efficiently and effectively, and self-organize the rule-based structure and control parameters “on the fly”.
 In other words, the SOF-PID control system is capable of “learning as you control”. Moreover, the SOF-PID control system 
can adapt to an entirely new operating environment effectively without the requirement of full retraining. \par

The remainder of this paper is organized as follows. Section II presents the general architecture of the SOF-PID control system. The evolving mechanism of the proposed system is given in section III, 
and the computtaional complexity analysis is given in section IV. Section V demonstrates the effectiveness and validity of the
 SOF-PID control system through numerical experiments with mobile robots in simulated and real-world environments, and Section VI concludes the paper.\par

\section{The Proposed SOC-PID Controller}
\subsection{General Architecture}
\begin{figure}[htbp]
\centerline{\includegraphics[height = 5.5cm, width= 8.2cm]{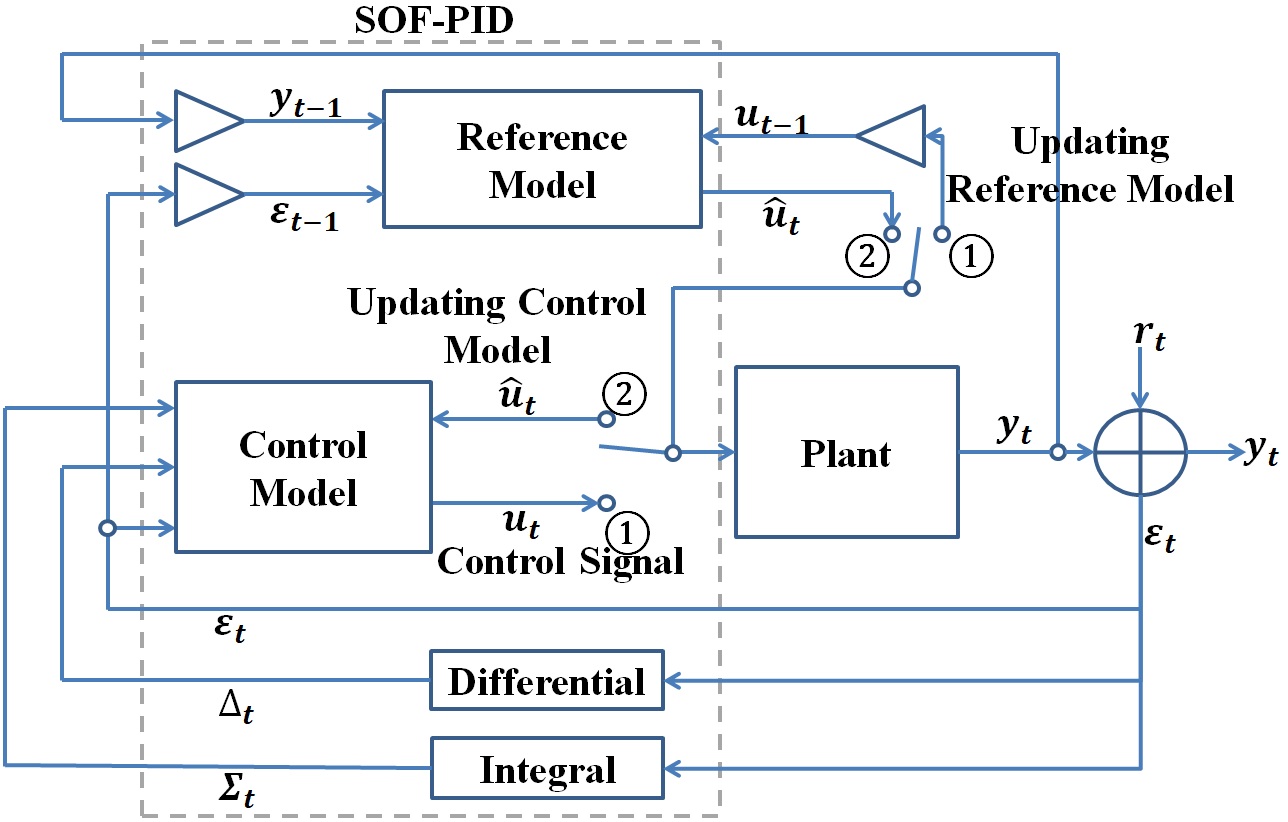}}
\caption{Architecture of the SOF-PID control system.}
\label{fig1}
\end{figure}

The proposed SOF-PID control system works as an inverse plant approximation model \cite{Angelov2004}, and the architecture is given in Fig. \ref{fig1}.  \par

As it is given by Fig. \ref{fig1}, SOF-PID control system consists of a control model and a reference model. The control model is for controlling the plant, and the reference model provides the desired output for the control model. Each models has an IF…THEN rule-based structure and is implemented using the first-order autonomous learning multi-model (ALMMo) neuro-fuzzy system \cite{Angelov2018} as the “learning engine”.  The proposed SOF-PID control system also offers the possibility of implementing several subsets of PID-based controllers, e.g., P, PI, PD, etc. In this paper, we consider the general implementation of PID-based controllers by taking all three (proportional, integral, derivative) components as the control model inputs.\par

For the control model, the ALMMo system will self-organise during the process, a set of IF…THEN rules from the controller inputs $\bm{x}_t=[\epsilon_t,\Sigma_t,\Delta_t ]^T$, namely, the tracking error, $\epsilon_t$ (the difference between the plant output, $y_t$ and the desired trajectory, $r_t$), the discrete derivative and integral of the tracking error, $\Delta_t$ and $\Sigma_t$:
\begin{equation}\label{eq1}
\epsilon_t=r_t-y_t;~~\Delta_t =\epsilon_t-\epsilon_{t-1};~~\Sigma_t=\sum_{k=0}^{t-1}\epsilon_k;
\end{equation}
where $t$ denotes the current control step; $t-1$ and $t$ stand for the consecutive control steps.\par

Each IF…THEN rule is formulated in the following form as given below, and thus, it can be viewed as a PID controller in its THEN part, but with the parameters learned from the historical data and zone of influence limited and determined by its IF part:
\begin{equation}\label{eq2}
\mathbf{R}_i^c:~IF~(\bm{x}_t\sim\bm{p}_{i,t-1} )~THEN~(\bm{u}_{i,t}=\bm{a}_{i,t-1}^T \bar{\bm{x}}_t )                 
\end{equation}
where $\sim$ denotes the similarity or a fuzzy degree of satisfaction/membership \cite{Angelov2012b}; $\bar{\bm{x}}_t=[\bm{x}_t^T,1]^T$; $\bm{p}_{i,t-1}$ is the prototype of the $ith$ IF…THEN rule at the current step, which is identified and updated from the historical control inputs $\bm{x}_1$, $\bm{x}_2$,…, $\bm{x}_{t-1}$ during the control process; $\bm{a}_{i,t-1}=[P_{i,t-1},I_{i,t-1},D_{i,t-1},R_{i,t-1}]^T$; $P_{i,t-1}$, $I_{i,t-1}$ and $D_{i,t-1}$ are the controller gains; $R_{i,t-1}$ is the compensation of the operating point; $u_{i,t-1}=P_{i,t-1} \epsilon_t+I_{i,t-1}\Sigma_t+D_{i,t-1}) \Delta_t+R_{i,t-1}$ denotes the output of this IF…THEN rule. Therefore, each IF…THEN rule in the rule base of the control model can be viewed as a PID controller (with the structure given by Fig. ref{fig12}), and the overall control signal produced by the control model is formulated by equation (\ref{eq3}) as a fuzzily weighted linear combination of outputs of the identified IF…THEN rules \cite{Angelov2018}:
\begin{equation}\label{eq3}
u_t=f_c(\bm{x}_t)=\sum_{i=1}^{M_c}\lambda_{i,t}u_{i,t}.
\end{equation}
where $f_c (\cdot)$ is the mathematical model between the inputs and output of the control model, which itself is a non-linear function approximated by the ALMMo neuro-fuzzy system ; $M_c$ is the number of IF…THEN rules in the rule base of the control model; $\lambda_{i,t}$ denotes the firing strength of the $ith$ IF…THEN rule, which is calculated as relative data density determined by the following expression \cite{Angelov2018}:
\begin{equation}\label{eq4}
\lambda_{i,t}=\frac{\gamma_{i,t}(\bm{x}_t)}{\sum_{k=1}^{M_c}\gamma_{k,t}(\bm{x}_t)};
\end{equation}
where $\gamma_{k,t}$ denotes the local data density calculated based on the data cloud, $\mathbf{C}_k^c$ associated with the prototype of the $kth$ IF…THEN rule \cite{Angelov2018}:
\begin{equation}\label{eq5}
\begin{split}
&\gamma_{k,t}(\bm{x}_t)\\
&=\frac{1}{1+\frac{S^2_{k,t-1}||\bm{x}_{t}-\bm{p}_{k,t-1}||^2}{(S_{k,t-1}+1)(S_{k,t-1}\chi_{k,t-1}+||\bm{x}_t||^2)-||\bm{x}_t-S_{k,t-1}\bm{p}_{k,t-1}||^2}};
\end{split}
\end{equation}
where $\mathbf{C}_k^c$  is composed of the historical controller inputs associated with prototype $\bm{p}_{k,t-1}$; $\bm{p}_{k,t-1}$ is the support of  $\mathbf{C}_k^c$ , namely, the number of members; $\chi_{k,t-1}$ is the average scalar product, which is calculated by $\chi_{k,t-1}=\frac{1}{S_{k,t-1}}\sum_{\bm{x}\in\mathbf{C}_k^c}||\bm{x}||^2$; $||\bm{x}||$ denotes the norm of $\bm{x}$ and can be calculated by $||\bm{x}||=\sqrt{\bm{x}^T\bm{x}}$.\par

The reference model shares the same operating mechanism as the control except for the difference in the IF…THEN rule bases. The IF…THEN rules identified by the reference model are formulated in the following form due to the different input signal:
\begin{equation}\label{eq6}
\mathbf{R}_i^r:~IF~(\bm{z}_t\sim\bm{q}_{i,t-1} )~~THEN~(\hat{u}_{i,t}=\bm{b}_{i,t-1}^T \bar{\bm{z}}_t)                 
\end{equation}
where $\bm{z}_t=[\epsilon_(t-1),y_(t-1) ]^T$; $\bar{\bm{z}}_t=[\bm{z}_t^T,1]^T$; $\bm{q}_{i,t-1}$ is the prototype of the $ith$ IF…THEN rule of the reference model, and it is identified and updated from the historical control inputs $\bm{z}_1$, $\bm{z}_2$,…, $\bm{z}_{t-1}$ during the control process; $\bm{b}_{i,t-1}=[Q_{i,t-1},Y_{i,t-1},W_{i,t-1}]^T$; $Q_{i,t-1},$, $Y_{i,t-1}$ are the controller gains; $W_{i,t-1}$ is the compensation; $\hat{u}_{i,t}=Q_{i,t-1} \epsilon_{t-1}+Y_{i,t-1} y_{t-1}+W_{i,t-1}$. The mathematical model between the inputs and output of the reference model is denoted as: $\hat{u}_{i,t}=f_r (\bm{z}_t)$.\par

In the following subsection, the operating mechanism of the SOF-PID control system is described in detail.

\begin{figure}[htbp]
\centerline{\includegraphics[height = 4.2cm, width= 8.2cm]{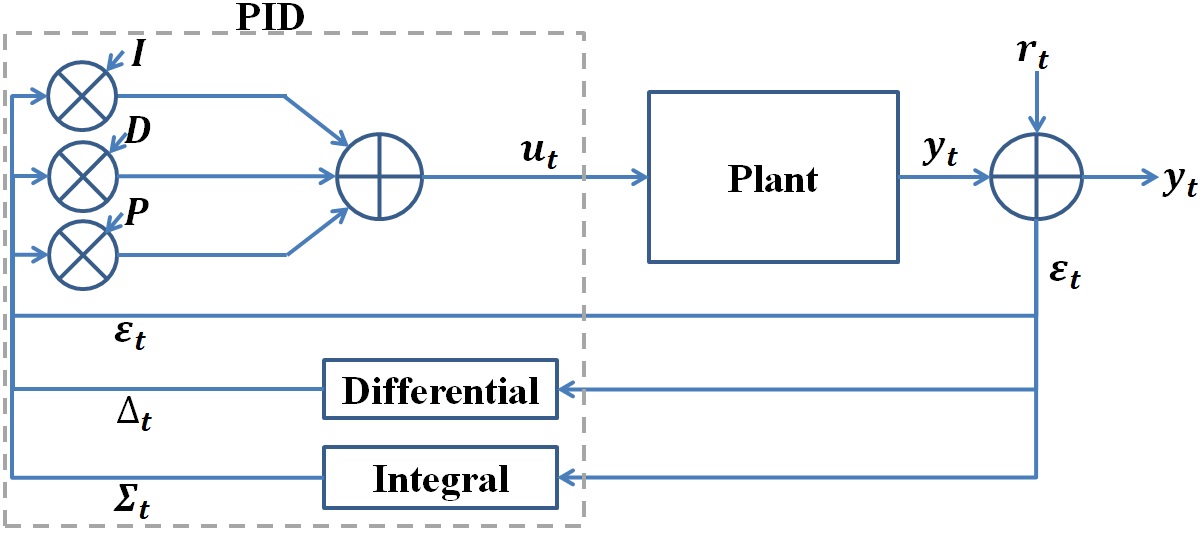}}
\caption{Architecture of a PID controller.}
\label{fig2}
\end{figure}
\subsection{Operating Mechanism}
The SOF-PID control system requires to be initialized because the reference model of the proposed system is designed to provide the desired output for the control model. Therefore, in the first $N$ control steps, the SOF-PID control system uses a PID controller with the same architecture as depicted in Fig. \ref{fig2} to control the plant, meanwhile it keeps collecting the data for initialization. In this paper, we use the PID controller because of its simplicity and ease of implementation. Nonetheless, one can also initialize the system by legacy data or using controllers of other types, i.e. fuzzy controllers, self-evolving controllers for the starting period.\par
\textbf{Stage 0}. For the first $N$ control steps, the proposed system uses a PID controller to control the plant; meanwhile, keeps collecting the data for initialization.\par
\textbf{Stage 1}. At the control step $t=N$, both the control and reference models are initialized based on the historical data, and then, it goes to \textbf{Stage 2}. \par
\textbf{Stage 2}. The  proposed system reads the current tracking error ($t\leftarrow t+1$) and calculates the discrete derivative and integral of the tracking error, namely, $\epsilon_t$, $\Sigma_t$ and  $\Delta_t$ using equation (\ref{eq1}).\par
\textbf{Stage 3}. The reference model produces the desired output $\hat{u}_t$ based on the inputs: $\bm{z}_t=[\epsilon_{t-1},y_{t-1} ]^T$ and updates its structure and meta-parameters based on the control signal of the previous control step $u_{t-1}$.\par
\textbf{Stage 4}. The control model produces the control signal $u_t$ based on the inputs: $\bm{x}_t=[\epsilon_t,\Sigma_t,\Delta_t ]^T$ and updates its structure and meta-parameters based on the desired output, $\hat{u}_t$.\par
\textbf{Stage 5}. The proposed system goes back to \textbf{Stage 2} for the next time control step.\par
It has to be stressed that the amount of historical data for initializing the SOF-PID control system, namely, $N$ is not a problem- and user-specific parameter, and it can be determined based on the user preference. Generally, the larger $N$ is, the better and more stable the proposed system performs. Meanwhile, in the numerical experiment part (section VI), we will demonstrate that the SOF-PID control system surpasses the alternatives with a small value of $N$.

\section{Self-Evolving Mechanism of SOF-PID Control System}
In this section, the evolving mechanism of the proposed SOF-PID control system will be presented. As it has been stated in the previous sections, the SOF-PID control system is composed of a pair of control and reference models, and both models employ a first-order ALMMo system as the “learning engine”. Both models self-organize and self-evolve their system structure and meta-parameters following the same algorithmic procedure, and, thus, we will focus on describing the evolving mechanism of the control model. The evolving mechanism of the reference model follows the same principles. It is worth to be noticed that despite the SOF-PID control system requires to be initialized by historical data, the control and reference models learn from data on a sample-by-sample basis.\par
The brief algorithmic procedure of the control model is given as follows \cite{Angelov2018}.\par
\textbf{\textit{Step 0. System Initialization:}}
The global meta-parameters of the control model are initialized by the input at the first control step, $\bm{x}_t=[\epsilon_t,\Sigma_t,\Delta_t ]^T$ ($t=1$):
\begin{equation}\label{eq7}
\bm{\mu}_t\leftarrow\bm{x}_t;~~X_t\leftarrow||\bm{x}_t||^2;~~M_c\leftarrow 1;
\end{equation}
where $\bm{\mu}_t$ and $X_t$ are the global mean and average scalar product, respectively. \par

The first data cloud of the control model, $\mathbf{C}_{M_c}^c$ is initialized by $\bm{x}_t$: $\mathbf{C}_{M_c}^c\leftarrow\{\bm{x}_t\}$. The meta-parameters of $\mathbf{C}$, which includes the prototype, $\bm{p}_{M_c,t}$, scalar product, $X_{M_c,t}$, support, $S_{M_c,t}$, accumulated firing strength, $\Lambda_{M_c,t}$, utility, $\eta_{M_c,t}$,  and the control step when $\mathbf{C}_{M_c}^c$ is initialized, $I_{M_c}$, are given as:
\begin{equation}\label{eq8}
\begin{split}
&\bm{p}_{M_c,t}\leftarrow\bm{x}_t;~~\chi_{M_c,t}\leftarrow||\bm{x}_t||^2;~~S_{M_c,t}\leftarrow1;\\
&\Lambda_{M_c,t}\leftarrow1;~~\eta_{M_c,t}\leftarrow1;~~I_{M_c}\leftarrow1;
\end{split}
\end{equation}
The first IF…THEN rule in the rule base of the control model is initialized as:\par
\begin{equation}\label{eq9}
\mathbf{R}_{M_c}^c:~IF~(\bm{x}_t\sim\bm{p}_{M_c,t} )~THEN~(\bm{u}_{M_c,t}=\bm{a}_{M_c,t-1}^T \bar{\bm{x}}_t )                 
\end{equation}
and the consequent parameters of the IF…THEN rule are initialized as:\par
\begin{equation}\label{eq10}
\bm{a}_{M_c,t}\leftarrow\mathbf{0}_{(L+1)\times1};~~\mathbf{\Theta}_{M_c,t}\leftarrow\Omega_o\mathbf{I}_{(L+1)\times(L+1)};
\end{equation}
where $L$ is the number of inputs, $L=3$ for the control model and $L=2$ for the reference model; $\mathbf{0}_{(L+1)\times1}$ is a $(L+1)\times1$ dimensional vector; $\mathbf{\Theta}_{M_c,t}$ is the co-variance matrix of the IF…THEN rule $\mathbf{R}_{M_c}^c$; $\mathbf{I}_{(L+1)\times(L+1)}$ is a $(L+1)\times(L+1)$ identity matrix; $\Omega_o$ is a user-controlled parameter initializing the co-variance matrix and $\Omega_o=10$, which follows the same setting as \cite{Angelov2018}. \par
\textbf{\textit{Step 1. Control Signal Generation:}} For the next control step ($t\leftarrow t+1$), the new input $\bm{x}_t=[\epsilon_t,\Sigma_t,\Delta_t ]^T$ is available. Each IF…THEN rule within the control model produces the firing strength $\lambda_{i,t}$ ($i=1,2,…,M_c$) using equation (\ref{eq4}). Then, the control signal is generated by: $u_t=f_c (\bm{x}_t )$.\par
\textbf{\textit{Step 2. Global Meta-Parameter Updating:}} The global mean and average scalar product are updated by $\bm{x}_t$ using the following equations, respectively:
\begin{equation}\label{eq12}
\bm{\mu}_t\leftarrow\frac{t-1}{t}\bm{\mu}_{t-1}+\frac{1}{t}\bm{x}_t;~~X_t\leftarrow\frac{t-1}{t}X^2_{t-1}+\frac{1}{t}||\bm{x}_t||^2;
\end{equation}\par
The data densities at $\bm{x}_t$ and the identified prototypes $\bm{p}_{i,t-1}$ ($i=1,2,…,M_c$) are, then, calculated by:
\begin{equation}\label{eq14}
\gamma_t(\bm{w})=\frac{1}{1+\frac{||\bm{w}-\bm{\mu}_t||^2}{X_t-||\bm{\mu}_t||^2}};
\end{equation}
where $\bm{w}=\bm{x}_t,\bm{p}_{1,t-1},\bm{p}_{2,t-1},…,\bm{p}_{M_c,t-1}$.

\textbf{\textit{Step 3. System Structure Updating:}} To update the system structure, \textbf{Condition 1} is firstly checked:
\begin{equation}\label{eq14}
\begin{split}
\textbf{Condition 1:~}& IF~(\gamma_t(\bm{x}_t)>\max_{i=1,2,...,M_c}(\gamma_t(\bm{p}_{i,t-1})))\\
& OR~(\gamma_t(\bm{x}_t)<\min_{i=1,2,...,M_c}(\gamma_t(\bm{p}_{i,t-1})))\\
& THEN~(\bm{x}_t~is~a~new~prototype)\\
\end{split}
\end{equation}
If\textbf{ Condition 1} is met, \textbf{Condition 2} is, then, checked to determining whether the new data cloud that is associated with $\bm{x}_t$  is overlapping with any of the previously identified data clouds:
\begin{equation}\label{eq15}
\begin{split}
\textbf{Condition 2:~}& IF~(\gamma_t(\bm{x}_t)>0.8)\\
& THEN~(\bm{x}_t~is~very~close~to~\bm{p}_{i,t-1})\\
\end{split}
\end{equation}
where $\gamma_t(\bm{x}_t)$i s calculated by equation (\ref{eq5}). 
If both \textbf{Conditions 1} and {2} are satisfied, the nearest data cloud $\textbf{C}_{n*}^c$ to $\bm{x}_t$ is found out by equation (\ref{eq16}):
\begin{equation}\label{eq16}
n*=\underset{i=1,2,...,M_c}{\mathrm{argmax}}(||\bm{x}_t-\bm{p}_{i,t-1}||);
\end{equation}
and the new data cloud is merged with $\mathbf{C}_{n*}^c$  ($\mathbf{C}_{n*}^c\leftarrow\mathbf{C}_{n*}^c+\bm{x}_t$):
\begin{equation}\label{eq17}
\begin{split}
&S_{n*,t}\leftarrow\frac{\lceil S_{n*,t-1}+1\rceil}{2};~\bm{p}_{n*,t}\leftarrow\frac{\bm{p}_{n*,t-1}+\bm{x}_t}{2};\\
&\chi_{n*,t}\leftarrow\frac{\chi_{n*,t-1}+||\bm{x}_t||^2}{2};
\end{split}
\end{equation}
where $\lceil\cdot\rceil$ denotes the operation of rolling up to the nearest integer.\par
If \textbf{Condition 1} is met, and \textbf{Condition 2} is unsatisfied, the new data cloud with $\bm{x}_t$ as the prototype is added to the system structure ($M_c\leftarrow M_c+1$): $\textbf{C}_{M_c}^c\leftarrow\{\bm{x}_t\}$ with the meta-parameters set by equation (\ref{eq8}).\par
The IF…THEN rule $\textbf{R}_{M_c}^c$ corresponds to $\textbf{C}_{M_c}^c$ is initialized in the same form as equation (\ref{eq9}) with the consequent parameters set as:
\begin{equation}\label{eq18}
\bm{a}_{M_c,t-1}\leftarrow\frac{1}{M_c-1}\sum_{j=1}^{M_c-1}\bm{a}_{j,t-1};~~\mathbf{\Theta}_{M_c,t}\leftarrow\Omega_o\mathbf{I}_{(L+1)\times(L+1)}.
\end{equation}

Otherwise, if both \textbf{Conditions 1} and \textbf{2} are not met, $\bm{x}_t$ is assigned to the nearest data cloud $\mathbf{C}_{n*}^c$ with the meta-parameters updated as:\par
\begin{equation}\label{eq19}
\begin{split}
&S_{n*,t}\leftarrow S_{n*,t-1}+1;\\
&\bm{p}_{n*,t}\leftarrow\frac{S_{n*,t-1}}{S_{n*,t}}\bm{p}_{n*,t-1}+\frac{1}{S_{n*,t}}\bm{x}_t;\\
&\chi_{n*,t}\leftarrow \frac{S_{n*,t-1}}{S_{n*,t}}\chi_{n*,t-1}+\frac{1}{S_{n*,t}}||\bm{x}_t||^2.\\
\end{split}
\end{equation}\par
For each data cloud $\textbf{C}_{n*}^c$ that does not receive new member at the current control step, its meta-parameters are set as:
\begin{equation}\label{eq20}
S_{i,t}\leftarrow S_{i,t-1};~~\bm{p}_{n*,t}\leftarrow\bm{p}_{n*,t-1};~~\chi_{n*,t}\leftarrow\chi_{n*,t-1}.
\end{equation}\par
\textbf{\textit{Step 4. IF…THEN Rule Base Quality Monitoring:}} The firing strength of each IF…THEN rule, $\lambda_{i,t}$ is calculated by equation (\ref{eq4}) and the accumulated firing strength is updated ($i=1,2,…,M_c$):
\begin{equation}\label{eq21}
\Lambda_{i,t}\leftarrow\Lambda_{i,t-1}+\lambda_{i,t}. 
\end{equation}\par
The utility of each rule is calculated by: 
\begin{equation}\label{eq22}
\eta_{i,t}\leftarrow\frac{1}{t-I_i}\Lambda_{i,t}.  
\end{equation}\par
Based on the updated utility, \textbf{Condition 3} is used for removing the data cloud and fuzzy IF…THEN rule that contributes little to the overall control signal \cite{Angelov2010,Angelov2012b}:
\begin{equation}\label{eq23}
\begin{split}
\textbf{Condition 3:~}& IF~(\eta_{i,t}>\eta_{o})\\
& THEN~(\mathbf{R}_{n*}^c~and~\mathbf{C}_{n*}^c~are~removed)\\
\end{split}
\end{equation}
where $\eta_o$ is another user-controlled parameter for monitoring the quality of IF…THEN rules, and $\eta_o=0.1$ following \cite{Angelov2018}.\par
If $\mathbf{R}_{n*}^c$ and $\mathbf{C}_j^c$ meet \textbf{Condition 3}, they are removed from the system structure and $M_c\leftarrow M_c-1$. \par
\textbf{\textit{Step 5. Consequent Part Updating:}} After the system structure updating, the consequent parameters of the IF…THEN rules in the rule base are updated by the FWRLS method as ($i=1,2,…,M_c$) \cite{Angelov2004}:
\begin{equation}\label{eq24}
\mathbf{\Theta}_{i,t}\leftarrow\mathbf{\Theta}_{i,t-1}-\frac{\lambda_{i,t}\mathbf{\Theta}_{i,t-1}\bar{\bm{x}}_t\bar{\bm{x}}_t^T\mathbf{\Theta}_{i,t-1}}{1+\lambda_{i,t}\bar{\bm{x}}_t^T\mathbf{\Theta}_{i,t-1}\bar{\bm{x}}_t}.
\end{equation}
\begin{equation}\label{eq25}
\bm{a}_{i,t}\leftarrow\bm{a}_{i,t-1}+\lambda_{i,t}\mathbf{\Theta}_{i,t}\bar{\bm{x}}_t(\hat{u}_t-\bm{a}_{i,t-1}^T\bar{\bm{x}}_t)
\end{equation}

\textbf{\textit{Step 6. Rule Base Updating:}} All the IF…THEN rules in the rule base are updated with the new premise and consequent parameters ($i=1,2,…,M_c$):
\begin{equation}\label{eq26}
\mathbf{R}_{i}^c:~IF~(\bm{x}_t\sim\bm{p}_{i,t} )~THEN~(\bm{u}_{i,t}=\bm{a}_{i,t-1}^T \bar{\bm{x}}_t )        
\end{equation}
After this, the system goes back to \textbf{Step 1} for the next control step.\par

The reference model can be learned using the same principles as described in this section. It needs to be noticed that the inputs of the reference model are $z_t=[\epsilon_{t-1},y_{t-1}) ]^T$ and the desired output of the reference model is the output of the control model, $u_t$.
 Detailed mathematical proof for the uniform stability of the ALMMo system can be found in \cite{Angelov2018a}.

 \section{Experimental Study}
\begin{figure}[htbp]
\centerline{\includegraphics[height = 6cm, width= 6cm]{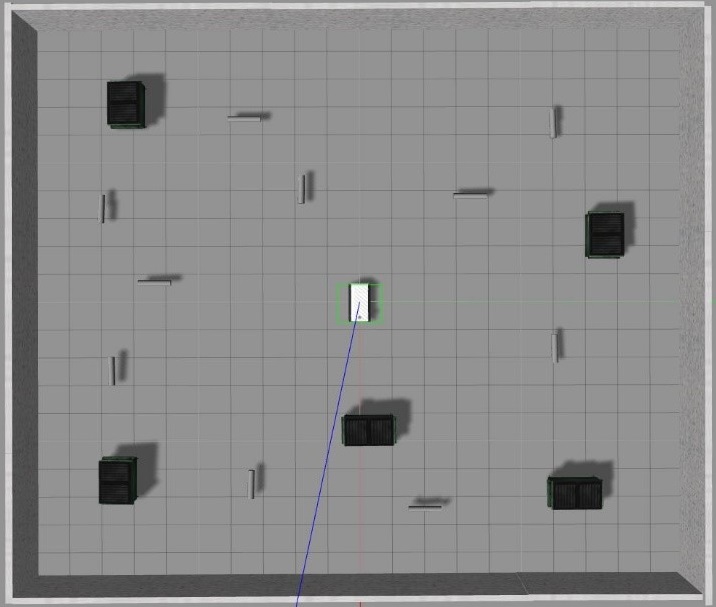}}
\caption{Simulated Gazebo environment with the mobile robot in the middle.}
\label{fig3}
\end{figure}

\begin{figure}[htbp]
\centerline{\includegraphics[height = 4cm, width= 3cm]{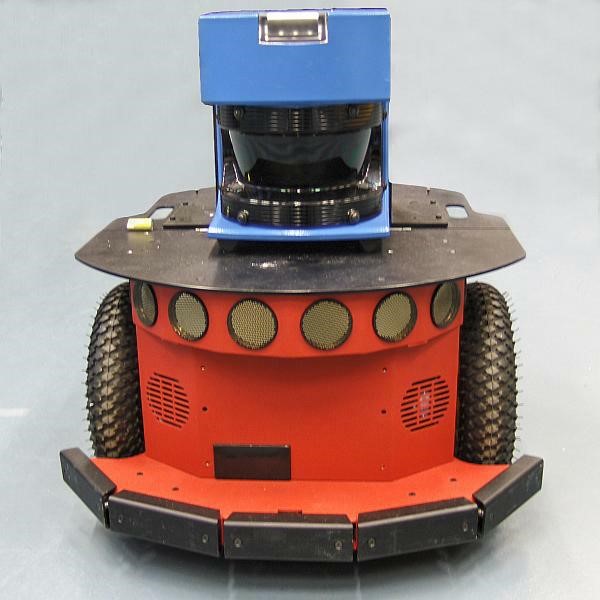}}
\caption{Pioneer 3DX robot with a laser finder on top and 16 sonar sensors; 8 at front and 8 at rear.}
\label{fig4}
\end{figure}

The performance of the proposed self-organizing fuzzy PID (SOF-PID) control system has been evaluated through several experiments. The experiments presented in this paper are organized as follows. Firstly, experiments in a simulated environment created in Gazebo (see Fig. \ref{fig3}) were performed to verify the validity and effectiveness of the proposed SOF-PID system. Then, the performance of the proposed controller was evaluated in various real-world environments using a Pioneer 3DX Mobile Robot (see Fig. \ref{fig4}).
\subsection{Computational Complexity Analysis}
In this section, the computational complexity of the proposed self-organizing fuzzy PID (SOF-PID) control system is analyzed. As the proposed control system is composed of a pair of control and reference models and the two components are, practically, the same except for the differences in terms of model inputs and output, we only analyze the computational complexity of the control model for clarity. It has to be stressed that the same conclusion applies to the reference model as well.\par
At the initial stage, the proposed approach requires to be primed by a PID controller until sufficient historical data has been collected. The computational complexity of the SOF-PID system at this stage is the same as a PID controller, namely, $O(LN)$. \par
Then, both control and reference models of SOF-PID are initialized by historical data, and continue to learn and produce control/reference signal at each control step. However, as the computational complexity of the control model is dynamically changing all the time, we assume that the analysis is conducted at the$tth$ time instance.\par
For the control model, the computational complexity of step 0 is negligible since this step concerns mainly parameter-initialization and will be performed once only. For step 1, the computational complexity of calculating the local data density-based firing strength is $O(LM_c)$ thanks to the recursive calculation form of equation (\ref{eq5}) and the complexity for calculating the control signal is also $O(LM_c)$.\par
Steps 2 and 3 concern mainly the system structure and meta-parameter updating. The complexity for calculating the data density in step 2 is $O(L(M_c+1))$. The computational complexity for updating meta-parameters globally and locally per rule is $O(L)$. The complexity for adding new rules to the control model is negligible. Therefore, the computational complexity of steps 2 and 3 is $O(L(M_c+1))$.\par
Step 4 is mostly for calculating the local data density, and thus, the complexity is $O(LM_c)$. Updating the consequent parameters of IF…THEN rules consumes significantly more computational resources because the co-variance matrix needs to be updated at each control step and, thus, the computational complexity of step 5 is $O(L^2 M_c)$. Step 6 updates the IF…THEN rules in the rule base, and its computational complexity is negligible as well. \par
Therefore, the overall computational complexity of the control model of the SOF-PID is $O(L^2 M_c)$ for each control step. The computational complexity of the reference model can be derived following the same principles.\par

\subsection{Experimental Settings}
\begin{figure}[htbp]
\centerline{\includegraphics[height = 6cm, width= 7.6cm]{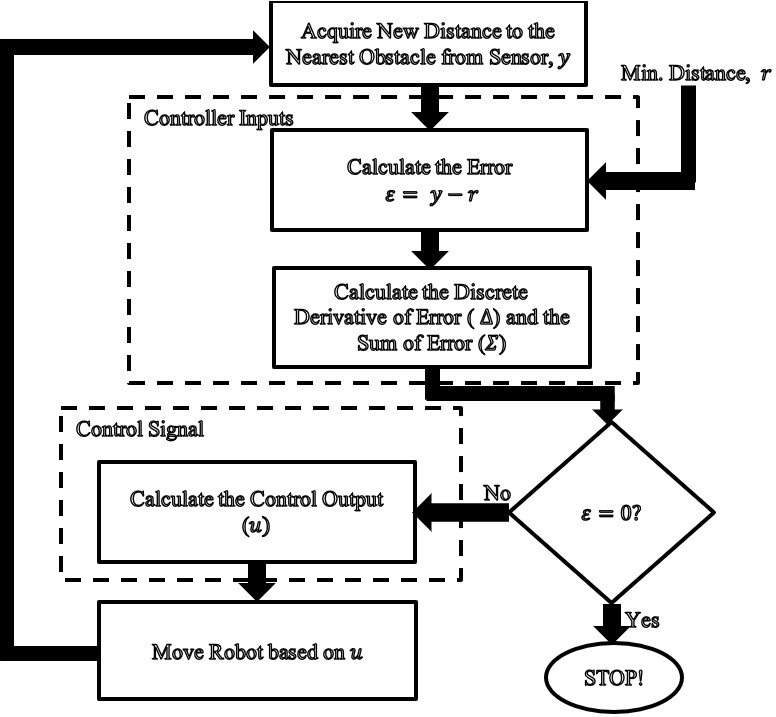}}
\caption{Block diagram of the SOF-PID controller for forward move.}
\label{fig5}
\end{figure}

In order to evaluate the performance of the proposed SOF-PID system, control of the forward motion of the mobile robot was implemented. The architecture of the overall forward move framework with a controller as a dotted block is shown in Fig. \ref{fig5}. \par
The simulated mobile robot (shown in the middle of Fig. \ref{fig3}) in the Gazebo environment is equipped with a $360^o$ scanning LIDAR sensor. The proposed SOF-PID system takes as the plant output, y the distance from the robot to the nearest detected object measured by the LIDAR senor. The Pioneer 3 robot is equipped with two types of input sensors that are a laser range finder and a set of 16-sonar sensors targeting at different angles, see Fig. \ref{fig4}. Similarly, the system takes as the plant output, y the distance between the robot and the nearest object in front measured by the two frontal sonars.  The minimum distance is set to $ r=1$ meter and $r=0.5$ meter for the two cases, respectively.
In the experiments performed in this paper, we use the same setting for the SOF-PID control system unless specifically declared otherwise. The number of time instances, N to initialize the SOF-PID using a PID prime controller is set as: $N=10$. The PID prime controller is defined as follows:
\begin{equation}
u=P\epsilon+I\Sigma+D\Delta
\end{equation}
where $P=0.25$, $D=0.1$ and $I=0$.\par
In order to justify the validity and effectiveness of the SOF-PID controller in various environments the experiments involve the commonly used PID \cite{Astrom1993} and Takagi-Sugeno (TS) fuzzy controllers\cite{Takagi1985} as comparative approaches following the same experimental setting. 
For a fair comparison, the PID controller used for comparison is the same as the PID prime controller. The TS fuzzy controller is defined as follows:
\begin{subequations}
\begin{equation}
IF~(\epsilon~is~Very~Low)~THEN (u=0.25\epsilon+0.001\Delta)
\end{equation}
\begin{equation}
\begin{split}
&IF~(\epsilon~is~Low)~AND~(\Delta~is~High)\\
&THEN (u=0.5\epsilon+0.002\Delta)\\
\end{split}
\end{equation}
\begin{equation}
\begin{split}
&IF~(\epsilon~is~Medium)~AND~(\Delta~is~High)\\
&THEN (u=0.5\epsilon+0.002\Delta)\\
\end{split}
\end{equation}
\begin{equation}
\begin{split}
&IF~(\epsilon~is~Medium)~AND~(\Delta~is~Low)\\
&THEN (u=\epsilon+0.03\Delta)\\
\end{split}
\end{equation}
\begin{equation}
IF~(\epsilon~is~High)~THEN (u=2\epsilon+0.02\Delta)
\end{equation}
\end{subequations}

\begin{figure}[htbp]
\centerline{\includegraphics[height = 6cm, width= 8cm]{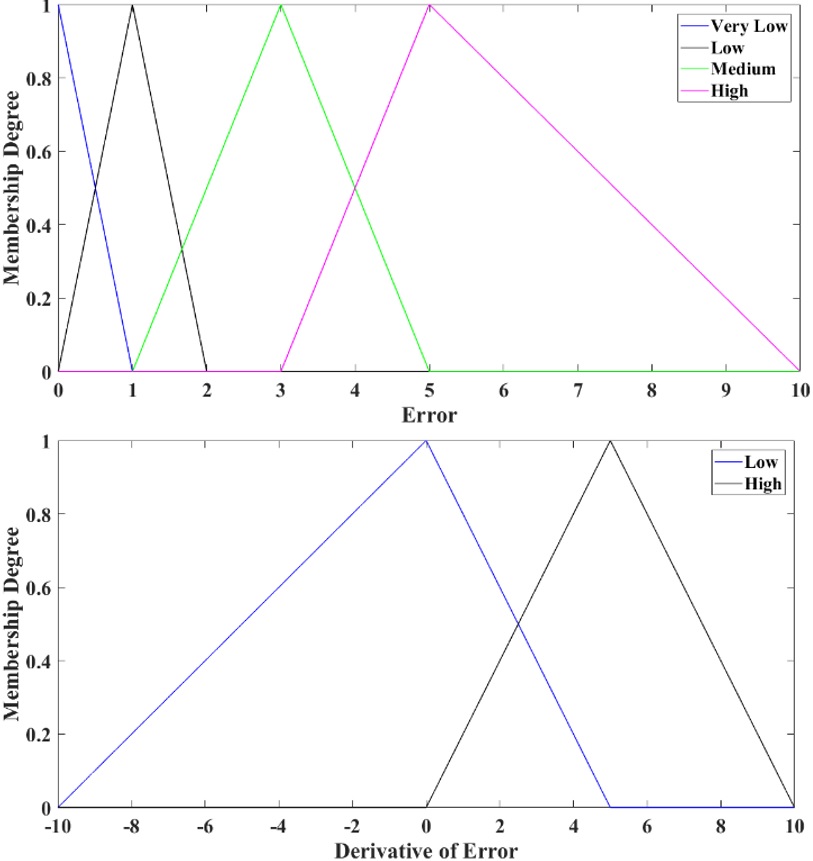}}
\caption{Membership functions of the antecedent part.}
\label{fig6}
\end{figure}

The membership functions of the antecedent parts of the fuzzy rules are of triangular type and visualized in Fig. \ref{fig6}.\par
It is worth to be noticed that due to the specific requirements of experimental scenarios, the mobile robot must not go across the target in case of potential damages it may cause, both PID prime controller and TS fuzzy controller do not involve the integral of tracking error as input to prevent the mobile robot from moving fast when it is close to the target. The mobile robot will stop immediately if it goes beyond the target. All the reported experimental results are the average of 10 Monte-Carlo experiments unless specially declared otherwise. \par

\subsection{Experimental Results}
In this subsection, the experiments are presented to verify the proposed concept and the general principles. \par
\begin{figure}[htbp]
\centerline{\includegraphics[height = 5cm, width= 7.5cm]{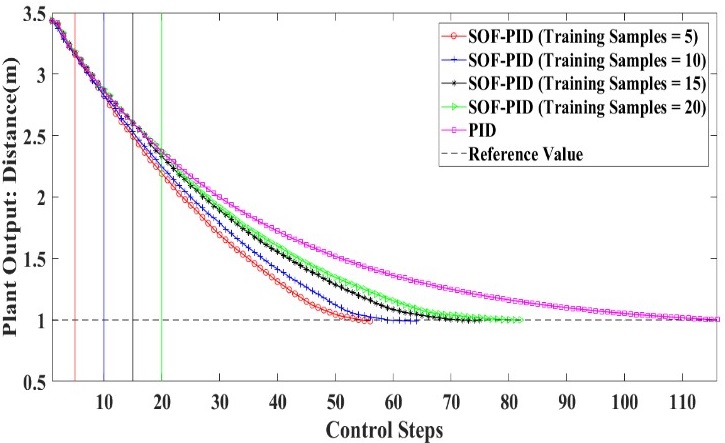}}
\caption{The influence of different $N$ on the SOF-PID control system performance in terms of plant output.}
\label{fig7}
\end{figure}
\begin{figure}[htbp]
\centerline{\includegraphics[height = 5cm, width= 7.5cm]{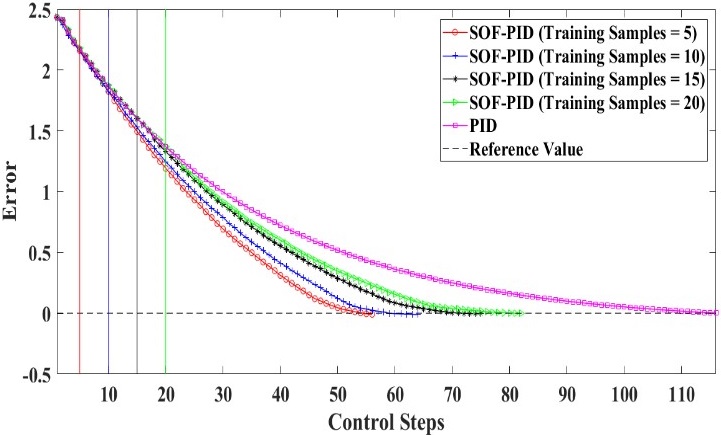}}
\caption{The influence of different $N$ on the SOF-PID control system performance in terms of tracking error.}
\label{fig8}
\end{figure}

Firstly, we perform experiments in the simulated Gazebo environment with the mobile robot to evaluate the influence of the different values of$ N=\{5,10,15,20\}$, namely, the number of steps used to initialize the controller, on the performance of the proposed SOF-PID controller. The correspondingly error and the plant output curves with different experimental settings are shown in Figs. \ref{fig7} and \ref{fig8}, respectively. The results of the PID controller (with same setting as the PID prime controller used by SOF-PID) are also presented as the baseline.

As we can see from both figures, the proposed SOF-PID requires a smaller number of steps to converge than a PID controller in all the cases. One may also notice that SOF-PID takes more control steps before the mobile robot reaches the target when we increase the prime-training control steps. This is due to the fact that the SOF-PID is essentially trying to approximate the control error converging process of the priming PID controller. As the integral of tracking error is not used as the input of the PID prime controller, the converging speed consistently decreases as the tracking error decreases. As a result, increasing the prime-training control steps will decrease the convergence speed of SOF-PID. Nonetheless, it has to be stressed that the SOF-PID control system is not mimicking the behavior of the PID prime controller, but is trying to learn the relationship between the controller inputs and desired controller output from the historical data collected by the PID prime controller and further to self-adjust its control gains to adapt to the changing environment. The curves in Figs. 7 and 8 further suggest that the newly proposed SOF-PID controller can adapt to the environment even if it is trained with only five samples. \par

\begin{figure}[htbp]
\centerline{\includegraphics[height = 8cm, width=7cm]{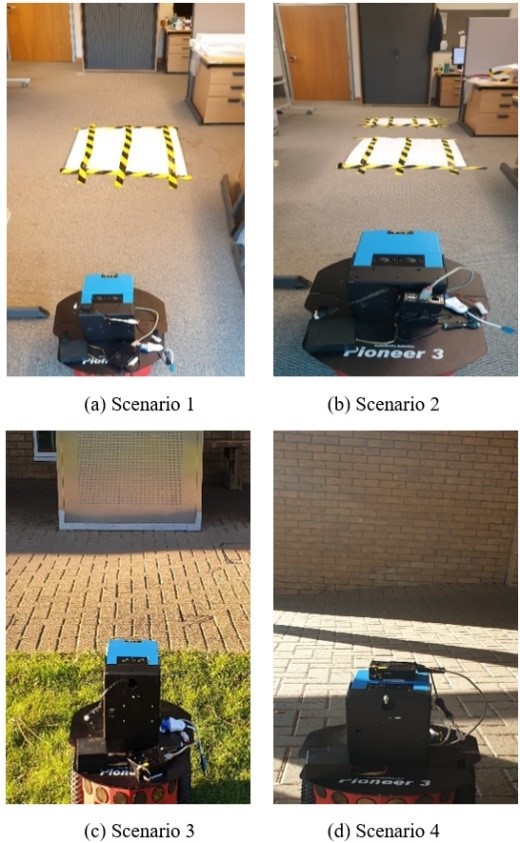}}
\caption{Four different experiment scenarios.}
\label{fig9}
\end{figure}

To further evaluate the performance of the proposed SOF-PID control system, four real-world experiments (see Figs. \ref{fig9}(a)-(d)) are performed. Details of the four experiment scenarios are given as follows:\par
Scenario 1 (Fig. \ref{fig9}(a)): The robot goes directly from one side of the room to a grey cupboard at the other side. The distance between the target (grey cupboard) and the starting position of the robot is 3.17 meters. There is one soft bump in the middle of the route. \par
Scenario 2 (Fig. \ref{fig9}(b)): This scenario is the same as the first one except that there are two soft bumps on the route. \par
Scenario 3 (Fig. \ref{fig9}(c)): In this scenario, the robot starts on the grass and goes straight to the target. The surface is changed to a brick road half way through. The distance between the starting position and the target is 4.07 meters. \par
Scenario 4 (Fig. \ref{fig9}(d)): In this scenario, the robot goes straight and stops before the wall. The robot, firstly, goes downwards on a slope, and then climbs another slope to reach the target. The distance between the starting position and the wall is 5 meters. \par

\begin{figure*}[htbp]
\centerline{\includegraphics[height =12.5cm, width=15cm]{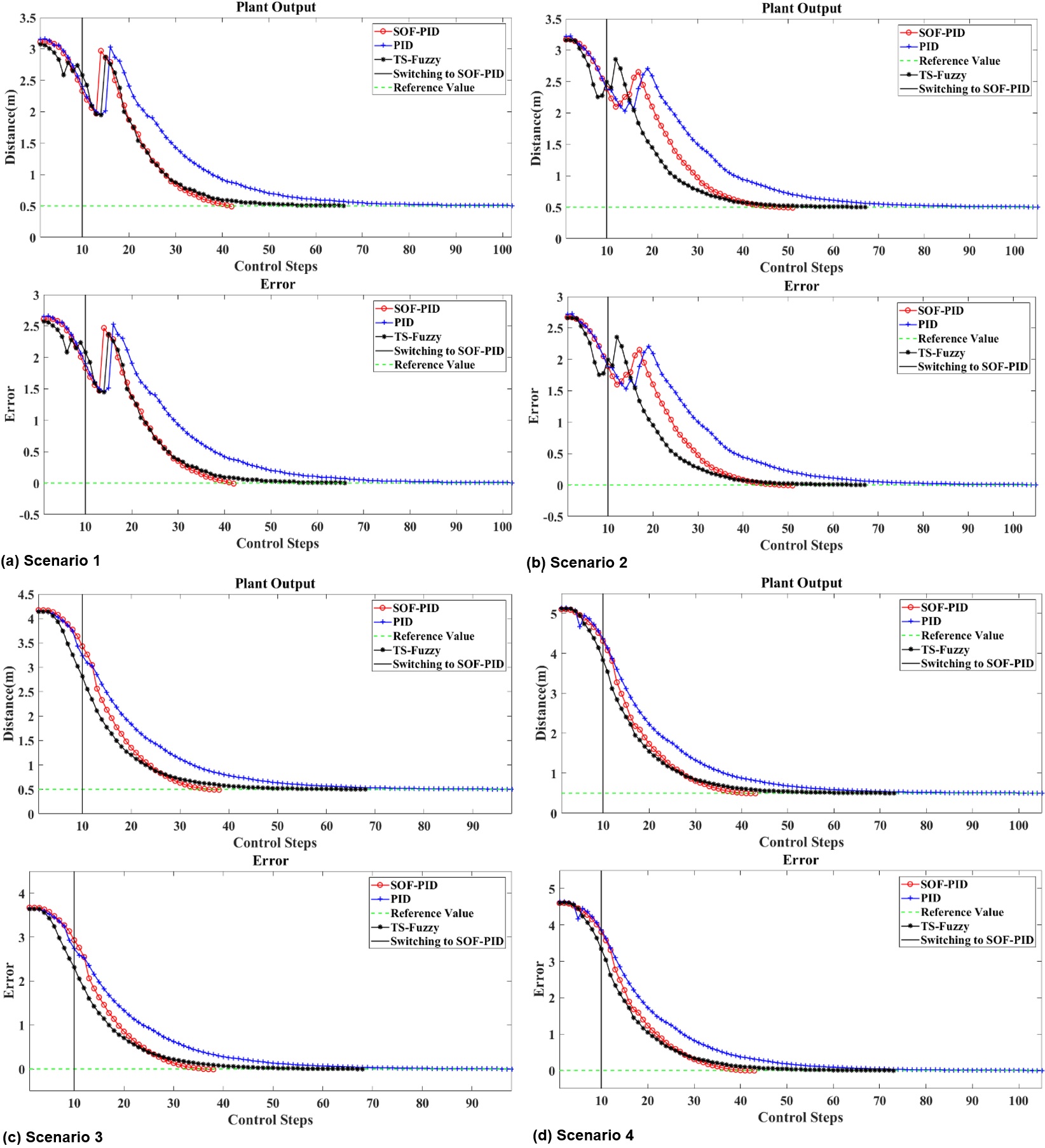}}
\caption{Performance comparison in different experiment scenarios.}
\label{fig10}
\end{figure*}

\begin{figure}[htbp]
\centerline{\includegraphics[height = 5.5cm, width= 7cm]{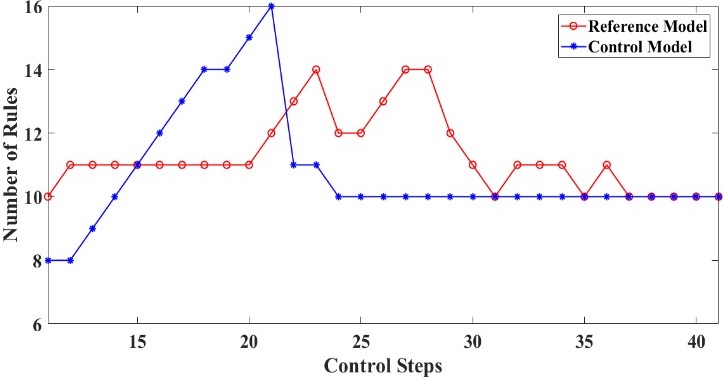}}
\caption{The evolution of the number of rules in the control and reference models during one of the experiments.}
\label{fig11}
\end{figure}

The ground surface of each experimental scenario is composed of either, patches with different levels of friction or slopes with different gradients. The mobile robot will go through a path with different surface conditions before reaching the target, which allow us to evaluate the adaptability of a controller to a changing environment.  \par
The experimental results obtained with the SOF-PID system are depicted in Figs. \ref{fig10}(a)-\ref{fig10}(d), and the performance is also compared with the performance of the PID and TS fuzzy controllers in terms of number of control steps for the robot to reach the target, the controller error and the plant output. From Fig. \ref{fig10} one can see that the SOF-PID control system performs better than the other controllers in all the scenarios. It takes much less control steps for the control error to converge to zero compared with the two alternatives. \par

In order to illustrate the transparency and human-interpretability of the proposed SOF-PID controllers, we also illustrate the evolution of the number of fuzzy IF…THEN rules in both, the control model and the reference model during a particular experiment conducted on scenario 1 in Fig. \ref{fig11}. It has to be noticed that Fig. \ref{fig11} starts from the $11th$ control step because the first 10 steps are performed with the PID controller used for priming. The IF…THEN rules of both models at the end of the process are also presented in Tables \ref{Ta1} and \ref{Ta2}, respectively.\par

\section{Conclusion}
In this paper, a novel SOF-PID control system is proposed, which consists of a pair of control and reference models. The control model and reference model are both implemented by first-order ALMMo neuro-fuzzy systems. Compared with the alternative controllers, the SOF-PID control system is capable of self-organizing and self-evolving its system structures and meta-parameters during the control process ``on the fly”, which enables the proposed system to adapt to new environments autonomously without a full re-training. We also mathematically prove the stability of the proposed system. The SOF-PID control system is a generic approach and offers different implementation possibilities. It effectively deals with nonlinear processes and requires no prior knowledge of the dynamics of the process and plant. It only requires a very short priming of 5-10 steps. Simulations on Gazebo platform and real-world experiments using Pioneer robot demonstrate that the proposed SOF-PID control system is able to provide stable control performance even in changing environments and its performance surpasses the alternative controllers. \par
It has to be stressed that the main purpose of this paper is to demonstrate the general concept and principles of SOF-PID, and provides primary experiments to prove the effectiveness and validity. As future works, we will further investigate the performance of the proposed SOF-PID control system in various types of challenging environments and compared with other advanced control approaches. We will also implement it for controlling other types of mobile robot movements, i.e. rotating, grabbing, as well as for other industrial processes, i.e. temperature and pressure control.\par

\begin{table*}[htbp]
\caption{THE IDENTIFIED IF…THEN RULES OF THE CONTROL MODEL DURING THE CONTROL PROCESS}
\begin{center}
\begin{tabular}{|c|c|}\hline
\#& \textbf{IF…THEN Rule}  \\ \hline
1 & $IF~(\bm{x}\sim[2.6500,4.3841,0.8815]^T )~THEN~(u=0.2171\epsilon-0.0017\Sigma+0.0076\Delta+0.0761)$\\ \hline
2 & $IF~(\bm{x}\sim[2.6300,5.5792,-0.0062]^T )~THEN~(u=0.2156\epsilon-0.0011\Sigma+0.0042\Delta+0.0651)$\\ \hline
3 &$IF~(\bm{x}\sim[2.5700,6.3352,-0.1987]^T )~THEN~(u=0.2139\epsilon-0.0007\Sigma+0.0025\Delta+0.0580)$\\ \hline
4& $IF~(\bm{x}\sim[2.5700,7.1291,0.0000]^T )~THEN~(u=0.2139\epsilon-0.0006\Sigma+0.0056\Delta+0.0577)$\\ \hline
5& $IF~(\bm{x}\sim[2.4700,7.8732,-0.3320]^T )~THEN~(u=0.2098\epsilon-0.0001\Sigma+0.0004\Delta+0.0494)$\\ \hline
6& $IF~(\bm{x}\sim[2.3600,8.5840,-0.3652]^T )~THEN~(u=0.2085\epsilon-0.0000\Sigma+0.0028\Delta+0.0488)$\\ \hline
7&$IF~(\bm{x}\sim[2.2200,9.2528,-0.4647]^T )~THEN~(u=0.2074\epsilon-0.0001\Sigma+0.0042\Delta+0.0469)$\\ \hline
8&$IF~(\bm{x}\sim[1.3400,18.1528,-0.2247]^T )~THEN~(u=0.2148\epsilon-0.0006\Sigma+0.0027\Delta+0.0585)$\\ \hline
9& $IF~(\bm{x}\sim[1.1600,18.6895,-0.3815]^T )~THEN~(u=0.2149\epsilon-0.0006\Sigma+0.0027\Delta+0.0586) $\\ \hline
10& $IF~(\bm{x}\sim[0.3395,21.2997,-0.1423]^T )~THEN~(u=0.2149\epsilon-0.0007\Sigma+0.0026\Delta+0.0587) $\\ \hline
\end{tabular}
\end{center}
\label{Ta1}
\end{table*}\par

\begin{table*}[htbp]
\caption{THE IDENTIFIED IF…THEN RULES OF THE CONTROL MODEL DURING THE CONTROL PROCESS}
\begin{center}
\begin{tabular}{|c|c|}\hline
\#& \textbf{IF…THEN Rule}  \\ \hline
1 & $IF~(\bm{z}\sim[2.6500,3.1500]^T )~THEN~(\hat{u}=0.1176\epsilon+0.1096y-0.0160)$\\ \hline
2 & $IF~(\bm{z}\sim[2.6300,3.1300]^T )~THEN~(\hat{u}=0.1154\epsilon+0.1082y-0.0144)$\\ \hline
3 &$IF~(\bm{z}\sim[2.4700,2.9700]^T )~THEN~(\hat{u}=0.1076\epsilon+0.1047y-0.0059)$\\ \hline
4& $IF~(\bm{z}\sim[0.2900,0.7900]^T )~THEN~(\hat{u}=0.1103\epsilon+0.1060y-0.0089)$\\ \hline
5& $IF~(\bm{z}\sim[0.2500,0.7500]^T )~THEN~(\hat{u}=0.1109\epsilon+0.1064y-0.0089)$\\ \hline
6& $IF~(\bm{z}\sim[0.2200,0.7200]^T )~THEN~(\hat{u}=0.1118\epsilon+0.1070y-0.0095)$\\ \hline
7&$IF~(\bm{z}\sim[0.1800,0.6800]^T )~THEN~(\hat{u}=0.1117\epsilon+0.1070y-0.0094)$\\ \hline
8&$IF~(\bm{z}\sim[0.1500,0.6500]^T )~THEN~(\hat{u}=0.1119\epsilon+0.1072y-0.0095)$\\ \hline
9& $IF~(\bm{z}\sim[0.1100,0.6100]^T )~THEN~(\hat{u}=0.1122\epsilon+0.1074y-0.0096) $\\ \hline
10& $IF~(\bm{z}\sim[0.0420,0.5420]^T )~THEN~(\hat{u}=0.1121\epsilon+0.1074y-0.0095)$\\ \hline
\end{tabular}
\end{center}
\label{Ta2}
\end{table*}\par

\end{document}